# Spin transfer and coherence in coupled quantum wells

M. Poggio, G. M. Steeves, R. C. Myers, N. P. Stern, A. C. Gossard, and D. D. Awschalom[*]

*Center for Spintronics and Quantum Computing, University of California, Santa Barbara, CA 93106*

**Abstract**

Spin dynamics of optically excited electrons confined in asymmetric coupled quantum wells are investigated through time resolved Faraday rotation experiments. The inter-well coupling is shown to depend on applied electric field and barrier thickness. We observe three coupling regimes: independent spin precession in isolated quantum wells, incoherent spin transfer between single-well states, and coherent spin transfer in a highly coupled system. Relative values of the inter-well tunneling time, the electron spin lifetime, and the Larmor precession period appear to govern this behavior.

The possibility of developing spin-based electronic devices has focused recent interest on the study of carrier spin dynamics in semiconductor nanostructures. In this vein, electrical control of electron spin precession and relaxation rates has been achieved in a number of quantum well systems[1,2,3]. The accessibility of spatially direct and indirect excitonic states with the application of an external electric field make coupled quantum well (CQW) systems[4] attractive for the study of electron spin dynamics. Extensive research has been devoted to indirect electron-hole pairs in CQWs[5,6,7,8] and to carrier tunneling between coupled wells[9,10,11]. Here, time resolved Faraday rotation (FR) experiments[12] on specifically engineered CQWs reveal the effect of inter-well tunneling on electron spin coherence. Since the electron g-factor depends strongly on quantum well width[13], electron spins in wells of unequal widths precess at different rates. When such wells are coupled through a tunneling barrier, spin precession rates are observed to either switch or tune continuously as a function of applied electric field.

The sample structure consists of a pair of undoped GaAs quantum wells (QWs) with $Al_{.33}Ga_{.67}As$ barriers grown by molecular beam epitaxy[14] on top of a low temperature $Al_{.33}Ga_{.67}As$ back gate structure[15] 1.3 μm from the surface. A Ni/Ge/Au/Ni/Au pad is annealed to contact the back gate, while a semi-transparent 1 mm² layer of Ti/Au deposited on the sample surface acts as the front gate. Applying a voltage $U_g$ across the gates creates a uniform electric field in the QWs up to 30 kV/cm with negligible leakage current (less than 50 μA). A positive value of $U_g$ corresponds to a positive voltage at the front gate with respect to the back gate. Different samples are grown with varying well widths $w$ and well separations $d$. Here we shall discuss 5 such samples: sample 7-2-10 consists of 10 nm QW grown on top of a 7 nm QW separated by a 2 nm barrier. Other samples include 7-6-10, 7-20-10, 8-4-8, and 5.7-3.8-7.7 using the same naming convention. Experiments are performed at 5 K in a magneto-optical cryostat with an applied magnetic field $B_0$ in the plane of the sample and with the laser excitation parallel to the growth direction.

Figs. 1a, b, and c show photoluminescence (PL) measurements as a function of $U_g$ and detection energy $E_d$ for samples 7-20-10, 7-6-10, and 7-2-10 respectively. Samples 7-20-10 and 7-6-10 in Figs. 1a and b show two distinct PL peaks each with FWHM of 2-3 meV centered around 1.54 and 1.57 eV corresponding to emission from the 10 nm and 7 nm wide wells respectively. The red-shift observed for each peak for $U_g < -2.0$ V agrees well with the Stark shift expected in QWs of similar thicknesses[16]. Fig. 1b, however, shows evidence of coupling between the two wells in the form of (i) a strongly Stark shifted indirect exciton peak appearing below $U_g = -2.0$ V, and (ii) a quenching of the higher energy PL peak together with an increase in the emission intensity of the lower energy peak around $U_g = 0.0$ V[5,17]. These features confirm that sample 7-6-10 with its 6 nm barrier between QWs is indeed a coupled system with a tunneling time $\tau$ shorter than the recombination lifetime $T_R$, while sample 7-20-10 with its much wider 20 nm barrier contains two uncoupled QWs with otherwise identical characteristics. Fig. 1c shows a single PL peak for sample 7-2-10 with a strong Stark shift at negative voltages indicating an even shorter value of $\tau$.

Time resolved FR measurements are performed in a magnetic field in order to examine carrier spin dynamics in CQWs. The measurement, which monitors small rotations in the linear polarization of laser light transmitted through a sample, is sensitive to the direction of spin polarization of electrons in the conduction band. By tuning the laser energy $E_L$ near the resonant absorption energy of different conduction band states, the polarization dynamics of these states can be selectively investigated. A 250 fs 76 MHz Ti:Sapphire laser produces pulses which are split into pump and probe with a FWHM of 8 meV and an average power of 2.0 mW and 100 µW respectively. The linearly polarized probe is modulated by an optical chopper at $f_1 = 940$ Hz and the circular polarization of the pump is varied by a photo-elastic modulator at $f_2 = 55$ kHz. Both beams are focused to an overlapping 50 µm spot on the semitransparent front-gate. Thus, polarized electron spins are injected and precess in a perpendicular field $B_0$. The time evolution of the spins is well described by the expression for FR as a function of pump-probe delay,

$$\theta_F(\Delta t) = \theta_\perp e^{-\frac{\Delta t}{T_2^*}} \cos(2\pi \nu_L \Delta t + \phi), \qquad (1)$$

where $\theta_\perp$ is proportional to the total spin injected perpendicular to the applied field, $T_2^*$ is the inhomogeneous transverse spin lifetime, $\Delta t$ is time delay between the pump and probe pulses, and $\phi$ is a phase offset. The Larmor frequency $\nu_L = g\mu_B B_0/h$ depends on the magnetic field $B_0$ and the Landé g-factor $g$ where $\mu_B$ is the Bohr magneton and $h$ is Planck's constant. It is important to note that our measurement is insensitive to hole spins due to their rapid spin relaxation (faster than 5 ps) in GaAs/Al$_x$Ga$_{1-x}$As QWs[18].

Fig. 2a shows FR measured in sample 7-6-10 at an applied magnetic field $B_0 = 6$ T as a function of both the $\Delta t$ and gate voltage $U_g$. Two distinct precession frequencies appear, as highlighted by the line-cuts at constant $U_g$ shown in Fig. 2b, with a sharp transition between the two occurring around $U_g = -2$ V, i.e. at the same voltage as the onset of the indirect excitonic peak in Fig. 1b. There is an accompanying 10-fold drop in the FR amplitude $\theta_\perp$ as a function of voltage.

The voltage dependent shift of $\nu_L$ in sample 7-6-10 is due to a change in the measured g-factor as shown in the inset to Fig. 2b. Here, the precession frequency, obtained by fitting (1) to data as shown in Fig. 2a, is plotted as a function of $B_0$ for two fixed voltages: $U_g = 0.0$ V and $U_g = -4.0$ V. The linear dependence of both distinct precession frequencies on $B_0$ demonstrates the presence of two independent g-factors ($|g| = 0.052$ +/- .001 and $|g| = 0.193$ +/- .005) whose relative weight can by controlled by $U_g$.

The dependence of the g-factor on $U_g$ is explored in greater detail in Fig. 3 for three samples with varying well separation $d$: 7-20-10, 7-6-10, and 7-2-10. FR data taken at $B_0 = 6$ T as a function of $\Delta t$ and $U_g$ (as shown in Fig. 2a) are Fourier transformed. Grayscale plots show the logarithm of the Fourier amplitude as a function of $U_g$ and of g-factor $|g|$ (extracted from the precession frequency $\nu_L$). Measurements are performed at a laser energy $E_L = 1.57$ eV resonant with the 7 nm well absorption. Fig. 3a shows the presence of the same two g-factors in sample 7-20-10, $|g| = 0.05$ and $|g| = 0.19$, as shown in Fig. 2. Experimental and theoretical literature confirms that these values of $g$ correspond to the 7 and 10 nm wide wells respectively[13]. Since $E_L$ is resonant with the 7 nm well absorption and detuned from the 10 nm well absorption by 20 meV, the Fourier amplitude of the $|g|$ = 0.05 oscillations is observed to be an order of magnitude larger than the $|g|$ = 0.19 oscillations, which correspond to the 10 nm well. Both g-factors show a weak dependence on $U_g$ corresponding to slightly increased penetration of the electron wave function into the barriers for $U_g <$ -2.0 V[19,20]. As shown schematically in the center and right panels of Fig. 3a, the inter-well tunneling time $\tau$ in this uncoupled sample is much longer than either the transverse spin lifetime $T_2$ or the recombination time $T_R$.

The effect of reducing $d$ to 6 nm and thus introducing inter-well coupling is shown in Fig. 3b. Here, a distinct switching behavior is observed between the 7 and 10 nm well g-factors as a function of $U_g$. Near $U_g$ = 0 V, spin polarized electrons are excited and detected in the 7 nm well, however, in contrast with the $d$ = 20 nm case, spin precession in the 10 nm well is not observed, even at a reduced amplitude. This behavior can be understood qualitatively from the center panel of Fig. 3b. Since the conduction band ground state of the 10 nm well is energetically lower then that of the 7 nm well, and because $d$ is sufficiently small that $\tau < T_R$, electrons tunnel from the 7 nm well into the 10 nm well. Assuming that $\tau$ is shorter than $T_2$ but longer than a spin precession period $1/\nu_L$, spin transfers incoherently. Because $\nu_L$ is unequal in the two wells, the incoherent tunneling randomizes the electron spin polarization in the 10 nm well, thereby destroying its spin coherence and quenching its FR signal. This picture is corroborated by the fact that in Fig. 1b, around $U_g$ = 0 V no significant PL is found from the 7 nm well while PL from the 10 nm well is increased, indicating that electrons excited in the 7 nm QW tunnel into the 10 nm QW before recombination. For $U_g < -2.0$ V, spin precession from the 7 nm well disappears and precession from the 10 nm well emerges. In this case, as shown in the right panel of Fig. 3b, the applied electric field has raised the 10 nm well ground state energy above the 7 nm ground state energy causing the incoherent tunneling to change directions. As a result, spin coherence in the 7 nm well is destroyed and its corresponding FR signal disappears. The amplitude of the 10 nm FR signal remains small due to the detuning of $E_L$. We can further conclude that near $U_g = -2.6$ V, where Fig. 1b shows that the electron ground state energy levels of the 10 nm and 7 nm wells are degenerate, incoherent tunneling occurs in both directions resulting in the destruction of spin coherence in both wells as shown in Fig. 3b.

Reduction of $d$ to 2 nm results in the smooth tuning of $g$ as a function of $U_g$ between the 10 and 7 nm values. In Fig. 3c, the g-factor is shown to change from $|g| = 0.19$ near $U_g = 0$ V to $|g| = 0.05$ for $U_g < 0$ V. As shown schematically in the right panels of Fig. 3c, this behavior corresponds to a system in which $\tau$ is shorter than $1/\nu_L$ resulting in electron spin wave functions which effectively span both quantum wells. As an electric field is applied across the structure, the relative amplitude of the wave function in each well is altered. Since the measured g-factor is a weighted average over the full electron wave function amplitude[1,21], $g$ is observed to tune continuously between the two single-well values. Near $U_g = 0$ V, the electron wave function amplitude is almost completely contained within the 10 nm well resulting in $|g| = 0.19$. For $U_g < 0$ V, $|g|$ approaches 0.05 as the wave function amplitude shifts to the 10 nm well.

In order to confirm the role of quantum well width and to rule out electron-hole exchange in causing the voltage dependence of the observed g-factor[22], experiments were done on two more structures. Fig. 4a, shows the Fourier transform of FR data taken at $B_0$ = 6 T and $E_L$ = 1.58 eV plotted as a function of |g| and $U_g$ (similar to grayscale plots in Fig. 3) for sample 8-4-8. The data indicate that spin oscillations occur at a single frequency corresponding to |g| = 0.105 +/- .005 with no measurable dependence on $U_g$. This g-factor corresponds to the expected value of g for an 8 nm wide GaAs QW. In addition, the lack of voltage dependence is expected in our model for a symmetric CQW structure; in particular, we find no evidence for a second excitonic g-factor. A similar Fourier transform is plotted in Fig. 4b for sample 5.7-3.8-7.7. Here we observe continuous tuning of g as a function of $U_g$ as seen in the highly coupled sample 7-2-10. In this case, |g| is observed to tune from 0.09 through 0 to 0.035 as $U_g$ is varied from +1.0 V to -2.0 V. Since GaAs/Al$_x$Ga$_{1-x}$As quantum wells are predicted to have negative g-factors for w greater than 6 nm and positive values of g for smaller values of $w$[13], we can conclude that this sample shows tuning of the g-factor from -0.09 through 0 to 0.035.

Experimental data taken from work by Snelling et al.[13] showing g as a function of w are plotted as crosses in Fig. 4c. A fit to their data is shown as a black line in order to guide the eye. Values of g extracted from FR data of our 5 samples and correlated to the well widths are plotted as filled circles in Fig. 4c. From our samples we obtain g-factors of 0.038, -0.052, -0.065, -0.105, and -0.193 for well widths of 5.74, 7, 7.65, 8, and 10 nm respectively. The sign of the g-factors was not explicitly measured, though an educated guess was made for the purposes of this plot. Fig. 4c shows close agreement of our g-factor data with previous measurements of *g* as a function of quantum well width.

The experimental data show electron spin precession in a fixed perpendicular magnetic field for CQW systems at low temperature. The effective g-factor of these structures is seen to depend on which well electrons occupy and on the strength of tunneling between wells. Spin-resolved measurements reveal two distinct regimes of inter-well coupling, resulting in either the abrupt switching or the continuous tuning of *g* as a function of an applied electric field. Since the width of each QW determines the g-factor of electrons confined therein, future CQW structures may be engineered to switch between a variety of precession rates, including positive and negative rates as observed in Fig. 4b and even *g* = 0. We thank F. Meier, Y. K. Kato, and A. Hollietner for many helpful discussions and acknowledge support from DARPA, ONR, and NSF. N. P. Stern acknowledges the Fannie and John Hertz Foundation.

**Figure Captions**

FIG. 1. PL intensity plotted on a logarithmic grayscale as a function of $U_g$ and $E_d$. A CW HeNe laser emitting at 1.96 eV is used to excite carriers at $B_0 = 0$ T. (a) PL from sample 7-20-10 shows two Stark shifted peaks corresponding to the 7 and 10 nm QWs without evidence of inter-well coupling. (b) PL from sample 7-6-10 (i) reveals a strongly Stark shifted indirect exciton peak, and (ii) shows the quenching of the 7 nm well PL peak and the corresponding greater intensity in the 10 nm well peak. (c) Sample 7-2-10 shows a single PL peak which is strongly Stark shifted.

FIG. 2. Dependence of time resolved FR data on $U_g$ and $B_0$ in sample 7-6-10. (a) FR plotted in a grayscale as a function of $U_g$ and $\Delta t$ for $B_0 = 6$ T and $E_L = 1.57$ eV. Note the appearance of only two precession frequencies and the sharp transition between the two. (b) Line cuts of the time resolved FR data shown in (a) for $U_g = -0.8$ V and $U_g = -4.0$ V. Inset: $\nu_L$ plotted as a function of $B_0$ for two gate voltages $U_g$. Data taken at $U_g = 0.0$ V is shown as crosses and data taken at $U_g = -4.0$ V is shown as filled circles. The solid lines are linear fits to the data.

FIG. 3. Dependence of g-factor on $U_g$ and $d$. (a) Fourier transform of time resolved FR data measured in sample 7-20-10 plotted on a logarithmic grayscale as a function of $|g|$ and $U_g$ at $B_0 = 6$ T and $E_L = 1.57$ eV. Note the presence of two g-factors with a weak dependence on $U_g$. Schematic band diagrams are shown in the middle and on the right for $U_g$ close to zero and for negative $U_g$ respectively. Electron spin is represented by blue

arrows, while holes are shown without spin to illustrate the rapid hole spin relaxation (less than 5 ps) in these systems. The thick red arrow indicates resonant excitation and detection of FR, while the thin dotted arrow refers to weaker, off-resonant FR. (b) Similar data is shown for sample 7-6-10 where switching between two g-factors is observed as a function of $U_g$. Panels on the right illustrate the destructive effect of incoherent tunneling on the spin coherence of the lower energy conduction electron state. (c) Continuous tuning of the g-factor is observed in sample 7-2-10 and the panels to the right schematically depict the electron ground states extending over both QWs.

FIG. 4. Dependence of g on QW width $w$. (a) Fourier transform of time resolved FR data measured in sample 8-4-8 plotted in a logarithmic grayscale as a function of $|g|$ and $U_g$ at $B_0 = 6$ T and $E_L = 1.58$ eV. Note that the g-factor, $|g| = 0.105$, has no observable dependence on $U_g$. (b) Similar data is plotted for sample 5.7-3.8-7.7 showing continuous tuning from $|g| = 0.09$, via $|g| = 0$ around $U_g = 0$ V, and to $|g| = 0.035$. The red line maps the peak position of the FFT in order to guide the eye. (c) $g$ is shown as a function of $w$. Data drawn from work by Snelling et al. is plotted as crosses and a fit to this data is shown as a solid line to guide the eye. Values of $g$ extracted from FR data of our 5 samples for different $w$ are plotted as filled circles.

**References**

[*] Address correspondence to: awsch@physics.ucsb.edu.

[1] G. Salis *et al*., Nature (London) **414**, 619 (2001).

[2] Y. Kato *et al*., Science **299**, 1201 (2003).

[3] O. Z. Karimov *et al*., Phys. Rev. Lett. **91**, 246601 (2003).

[4] R. Dingle, A. C. Gossard, and W. Wiegmann, Phys. Rev. Lett. **34**, 1327 (1975).

[5] Y. J. Chen, E. S. Koteles, B. S. Elman, and C. A. Armiento, Phys. Rev. B **36**, 4562 (1987).

[6] M. N. Islam *et al*., Appl. Phys. Lett. **50**, 1098 (1987).

[7] T. Fukuzawa, E. E. Mendez, and J. M. Hong, Phys. Rev. Lett. **64**, 3066 (1990).

[8] L. V. Butov, *et al*., Phys. Rev. Lett. **92**, 117404 (2004).

[9] A. Alexandrou *et al*., Phys. Rev. B **42**, 9225 (1990).

[10] H. G. Roskos *et al*., Phys. Rev. Lett. **68**, 2216 (1992).

[11] V. B. Timofeev *et al*., JETP Lett. **67**, 613 (1998).

[12] S. A. Crooker *et al*., Phys. Rev. B **56**, 7574 (1997).

[13] M. J. Snelling *et al*., Phys. Rev. B **44**, 11345 (1991).

[14] A. C. Gossard, IEEE J. Quant. Electr. **22**, 1649 (1986).

[15] K. D. Maranowski, J. P. Ibbetson, K. L. Campman, and A. C. Gossard, Appl. Phys. Lett. **66**, 3459 (1995).

[16] H.-J. Polland *et al*., Phys. Rev. Lett. **55**, 2610 (1985).

[17] T. B. Norris *et al*., Phys. Rev. B **40** 1392 (1989).

[18] T. C. Damen *et al*., Phys. Rev. Lett. **67**, 3432 (1991).

[19] E. L. Ivchenko, A. A. Kiselev, and M. Willander, Solid State Commun. **102**, 375 (1997).

[20] H. W. Jiang and E. Yablonovitch, Phys. Rev. B **64**, R41307 (2001).

[21] M. Poggio *et al.*, Phys. Rev. Lett. **91** 207602 (2003).

[22] I. Y. Gerlovin *et al.*, Phys. Rev. B **69**, 035329 (2004).

FIG. 1, M. Poggio et al.

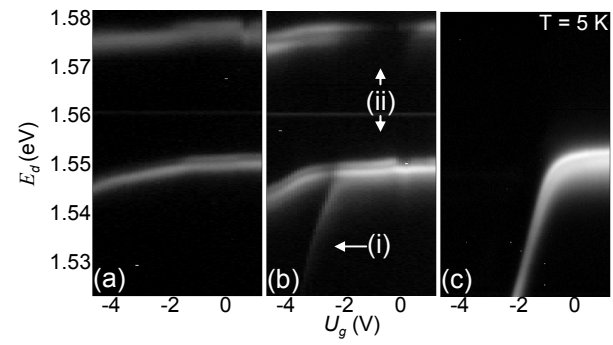

FIG. 2, M. Poggio et al.

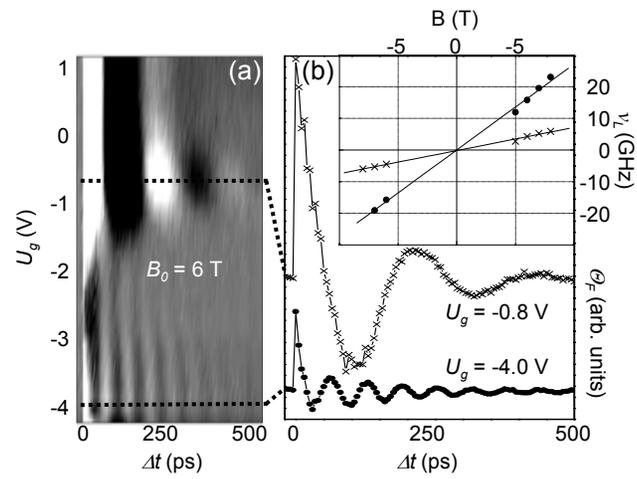

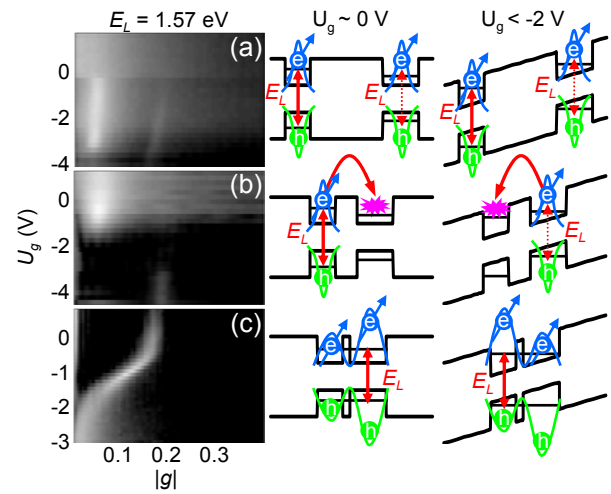

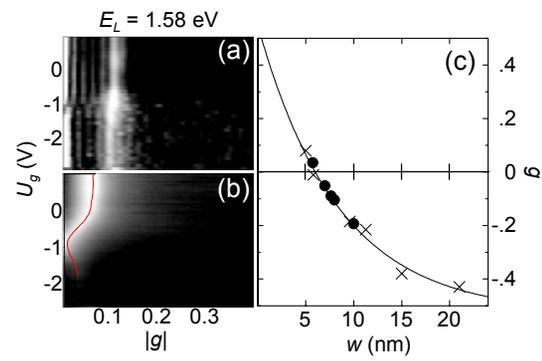